\documentclass[11pt]{article}
\pdfoutput=1
 \usepackage{mciteplus}
 \usepackage{tikz}
 \usepackage{color}
 \definecolor{darkblue}{rgb}{0.1,0.1,.7}
 \usepackage[colorlinks, linkcolor=darkblue, citecolor=darkblue, urlcolor=darkblue, linktocpage,hyperfootnotes=false]{hyperref} 
\usepackage{epsfig}
\usepackage{graphicx}
\usepackage{cite}
\usepackage{amsfonts}
\usepackage{amssymb}
\usepackage{bm}
\usepackage{latexsym}
\usepackage{amsmath}
\def\bq{\begin{quote}}
\def\eq{\end{quote}}
 at 10truept

\newcommand{\calo}{{\cal O}}
\newcommand{\calh}{{\cal H}}
\newcommand{\beq}{\begin{equation}}
\newcommand{\eeq}{\end{equation}}
\newcommand{\beqa}{\begin{eqnarray}}
\newcommand{\eeqa}{\end{eqnarray}}
\newcommand{\bea}{\begin{eqnarray}}
\newcommand{\eea}{\end{eqnarray}}

\def\roughly#1{\raise.3ex\hbox{$#1$\kern-.75em\lower1ex\hbox{$\sim$}}}

\begin{document}

\thispagestyle{empty}
\begin{titlepage}
  \bigskip

  \bigskip\bigskip

  \bigskip

\begin{center}
{\Large \bf {Black holes and other clues to the quantum structure of gravity}\footnote{Expanded version of a ``Letter of Interest" for Snowmass 2021.}}
    \bigskip
\bigskip
\end{center}

  \begin{center}

 \rm {Steven B. Giddings\footnote{\texttt{giddings@ucsb.edu}} }
  \bigskip \rm
\bigskip

{Department of Physics, University of California, Santa Barbara, CA 93106, USA}  \\
\rm

  \bigskip \rm
\bigskip
 
\rm

\bigskip
\bigskip

  \end{center}

\vspace{3cm}
  \begin{abstract}
Bringing gravity into a quantum-mechanical framework is likely the most profound remaining problem in fundamental physics. The ``unitarity crisis" for black hole evolution appears to be a key facet of this problem, whose resolution will provide important clues.  Investigating this
  raises the important structural question of how to think about subsystems and localization of information in quantum gravity.  Paralleling field theory, the answer to this is expected to be an important ingredient in the mathematical structure of the theory.  Perturbative gravity results indicate a structure different from that of quantum field theory, but suggest an avenue to defining subsystems.  If black holes do behave similarly to familiar subsystems, unitarity demands new interactions that transfer entanglement from them.  Such interactions can be parameterized in an effective approach, without directly addressing the question of the fundamental dynamics, whether that is associated with quantum spacetime, wormholes, or something else.  
  Since such interactions need to extend outside the horizon, that raises the question of whether they can be constrained, or might be observed, by new electromagnetic or gravitational wave observations of strong gravity regions.  This note overviews and provides connections between these developments.  

 \medskip
  \noindent
  \end{abstract}
\bigskip \bigskip \bigskip 

  \end{titlepage}


Reconciling gravity with quantum principles is likely the most profound remaining theoretical problem coming into this century.  This is expected to involve a description of quantum spacetime, and its resolution should provide a basic foundation for the rest of physics.  It is important to seek any helpful clues towards solving this problem.  The community has increasingly appreciated that reconciling black hole evolution with quantum mechanics is likely to provide a critical clue,
 and another important clue may be the novel structure of subsystems in the gravitational context.  These questions are in fact related.  This note will overview recent work on these questions, and in the process give further explanation of their connections, and will briefly discuss the possibility of observational constraints on or discovery of this new physics.

A large segment of the community believes that string theory, through the proposed AdS/CFT\cite{Maldacena:1997re} and related dualities, resolves the problem of quantum gravity.  However, if this is true, it seems that we do not yet know {\it how} string theory answers a number of central questions, such as those of defining localized observables, and describing cosmological evolution and that of black holes.  In particular, there is a question of how to define the ``holographic map" between bulk and boundary theories.  Examination of this question suggests that definition of this map ultimately requires solution of the nonperturbative completion of the bulk gravitational constraints\cite{SGHolo}, which appears tantamount to solving the problem AdS/CFT was supposed to solve.  Whatever the answer, anti de Sitter space does provide an important testing ground for quantum gravity questions and ideas.

{\bf The unitarity crisis.}  The problem of reconciling the existence of black holes with quantum mechanics appears to be a ``key" problem for quantum gravity, plausibly playing a role like explaining the atom did for quantum mechanics.  Our current description, based on local quantum field theory (LQFT) evolving on a semiclassical geometry, produces a crisis in  physics, commonly called the black hole information problem.  Specifically, if we can to good approximation describe a black hole (BH) and its environment as quantum subsystems of a bigger system, LQFT through Hawking's calculation\cite{Hawking:1974sw} implies that the BH builds up entanglement with its environment.  Locality of QFT implies that this entanglement cannot transfer from the BH subsystem.  If the BH disappears at the end of evolution, there is no longer a system to entangle with, and unitarity is violated.  But, failure to disappear, {\it e.g.} by leaving a remnant, leads to other paradoxical behavior\cite{Giddings:1994qt,Susskind:1995da}.  Worse still, violation of unitarity is apparently associated with catastrophic violations of energy conservation\cite{Banks:1983by}.\footnote{While this is a broad view of the community, a contrary viewpoint is that of \cite{Unruh:1995gn}.}

{\bf Mathematical structure of quantum gravity.}  The preceding description illustrates the connection to the important fundamental question:  how do we mathematically describe subsystems, and localization of information, in quantum gravity?  Indeed, various proposals for a resolution to the crisis rely on ideas that amount to challenging the view that a black hole is a quantum subsystem.  One of these is  the idea of soft quantum hair\cite{HPS1,HPS2,astrorev,astrorevisit,HHPS}, suggesting that information does not localize in a BH but instead is present in features of its exterior gravitational field.  Others include ideas associated with the proposal that entanglement generates spacetime connectedness, or ER=EPR\cite{vanR,MaSu}.  

It is worthwhile to compare to the question of localization of information in other quantum systems.  For finite or locally finite ({\it e.g.} lattice) systems, subsystems are described via factorization of the Hilbert space, $\calh = \calh_1\otimes\calh_2$ and information localizes in factors.  LQFT is more subtle, given the infinite entanglement existing between a region and its complement, associated with the von Neumann type III property.  Instead, as described in { \it e.g.} \cite{Haag}, subsystems can be associated with commuting subalgebras of the algebra of observables, {\it e.g.} field operators convolved with test functions with compact support in a given region.  

These subalgebras, describing subsystems, are in one-to-one correspondence with open sets of the background spacetime manifold.  They also have inclusion, overlap, {\it etc.} relations mirroring those of the open sets.  So, the structure defined on the Hilbert space by  this ``net" of subalgebras captures the topological structure of the manifold.  
This appears to be an important point, worth emphasizing:  from the quantum perspective, the topological structure of the underlying classical manifold is encoded in the Hilbert space description in this fashion.  

The property of commutativity also encodes the causal structure: commuting subalgebras correspond to spacelike separated regions.  
This is how the property of locality is hardwired into LQFT, which in flat space can be viewed as the answer to the question of how to reconcile the principles of quantum mechanics, the principles of relativity (Poincar\'e invariance), and the principle of locality.  

Importantly, gravity behaves differently.  First, there are no local gauge-invariant, or physical, observables\cite{Torre:1993fq}. One can  perturbatively construct gauge-invariant observables  that reduce to field theory observables in the weak-gravity limit, by ``gravitationally dressing" an underlying LQFT observable; the condition for gauge-invariance is that it commute with the GR constraints\cite{DoGi1}.  These observables are now nonlocal, and generically do not commute at spacelike separation.  Thus the basic locality property of LQFT is seen to be modified even at a leading perturbative level.

This raises the question of how to localize information, or define subsystems.  Given that the answer to this in LQFT incorporates the topological structure of the manifold and the key property of locality, one expects the answer to this question to be a key structural property in the mathematical description of quantum gravity.  In fact, giving a complete description of this mathematical structure may be the path to defining the quantum generalization of the spacetime manifold, needed in quantum gravity.

Perturbatively one can find a  structure that begins to describe such localization.  One way to see this is based on extending the notion of a splitting\cite{Doplicher:1984zz,Haag} in LQFT.  Given neighborhood $U$ and  $\epsilon$-extension of it $U_\epsilon$, one can find an embedding of a product of Hilbert spaces associated to $U$ and the complement $\bar U_\epsilon$ into the full Hilbert space, based on the ``split vacuum,"  giving a different kind of definition of ``subsystems."  Including gravity, the gravitational field of excitations in $U$ will extend into $\bar U_\epsilon$, making measurements outside $U$ depend on its internal excitations.  {\it However,} as shown in \cite{DoGi4,Giddings:2019hjc}  the gravitational field may be chosen in a ``standard" form so that measurements in   $\bar U_\epsilon$ only depend on the total Poincar\'e charges of the matter in $U$.  This indicates that one has a set of Hilbert space embeddings, labeled by these charges, and provides a candidate subsystem structure.

An  important problem for the future is to investigate this structure further, and improve our characterization of the localization of information, in terms of mathematical structure on Hilbert space.  The nonperturbative extension of gravitational dressing, and of this structure, is also closely connected to the question of how we might explain  holographic behavior of gravity\cite{SGHolo}

{\bf Quantum consistency for BHs.}  Given such a construction, and  an extension to BH backgrounds\cite{GiWe}, it appears that one can perturbatively describe localization of information, and, {\it e.g.}, seemingly  rule out the idea that the information in a BH is also present outside in its soft quantum hair.  And, if subsystems can be defined in this fashion in gravity, that returns us to the question of how BH evolution is unitary.  Suppose that a BH can be approximately described as a subsystem that localizes information.  If a BH can be described as a subsystem, and if evolution is required to be unitary, and if that subsystem ultimately disappears, then there is a ``theorem" that interactions must be present  that transfer entanglement from the BH to its environment.  

A question is to parameterize these interactions, and one can conservatively ask what such interactions 
 ``minimally" depart from the LQFT evolution, in an effective description.  It has been proposed in particular that if such interactions are ``soft," that is {\it e.g.} characterized by scales comparable to that of the BH rather than by microscopic scales, then they can have very limited effect on infalling observers, and thus preserve many of the essential features of BHs\cite{SGmodels,BHQIUE,Giddings:2012gc,SBGmodu,Giddings:2017mym}.  A contrasting possibility is that a BH is replaced by a new kind of ``hard" object, such as a firewall\cite{AMPS} or fuzzball\cite{Mathur:2008nj}.  In the former, soft, case, arguments based on preserving basic features of BH thermodynamics, and based on gedanken experiments of BH mining \cite{UnWamine,Lawrence:1993sg,Frolov:2000kx,Frolov:2002qd} suggest that the interactions couple universally to other fields, for example through a contribution to the effective hamiltonian
 \beq\label{deltah}
 \Delta H= \int dV_3 H^{\mu\nu}(x) T_{\mu\nu}(x)\ ,
 \eeq
 where $dV_3$ is the volume element on a time slice, $H^{\mu\nu}(x)$ are operators acting on the BH state, and $T_{\mu\nu}$ is the stress tensor (including perturbative gravitons).  More generally, and in the case of hard interactions, one may wish to consider other operators in the effective hamiltonian.
 
 {\bf Fundamental description?} 
 If interactions such as \eqref{deltah} give an {\it effective} description of interactions necessary to unitarize BH evolution, an important question is how they arise from a  {\it fundamental} description.  Possibly, they naturally arise in the appropriate dynamical evolution on the quantum spacetime once that is defined extending the ideas outlined above.  Conversely, the necessity of interactions such as \eqref{deltah}, implied by appropriate subsystem structure together with unitarity, appears to be furnishing an important clue regarding this dynamics.

{\bf Replica wormoles?}
Recent work has proposed that the underlying unitarizing dynamics is captured by {\it replica wormholes}\cite{Saad:2019lba,Penington:2019kki,Almheiri:2019qdq,MaMa1}.  Specifically, if the replica method is used to compute entropies, and a sum over euclidean geometries is carried out that extends the usual rules\cite{GiTu} of summing over configurations and taking traces to compute probabilities or entropies, one suggestively finds entropies that appear to follow the appropriate ``Page curve,"\cite{PageI,PageII}, with an entropy that declines in the latter phase of the BH evolution, as expected for decreasing entanglement of the BH and its radiation.

So far the underlying quantum amplitudes that are responsible for this behavior have not been clearly identified, raising the question of whether the argument simply arises from invention of clever rules that manage to reproduce the previously-known Page curve, or whether it rests on a more fundamental basis.  It has been suggested that replica wormholes are associated with spacetime wormholes and baby universe emission\cite{MaMa1,MaMa2}, extending work from the late 1980s\cite{Lavrelashvili:1987jg,Hawking:1987mz,GiStax,Cole,GiStinc,GiSt3Q}.  In the approximation of a weakly coupled ensemble of baby universes, these in the end lead to superselection sectors\cite{Cole,GiStinc}, called $\alpha$-vacua, in which evolution is of the standard LQFT form on scales large compared to the wormholes/baby universes, with effective coupling constants multiplying operators that describe baby universe absorption/emission.  Accordingly, if the wormholes are smaller than the BH, it is not clear how amplitudes are modified to avoid the original argument for loss of information in the new effective theory, with definite $\alpha$'s.  Possibly, baby universe sizes extending outside the horizon could play a role, but a clear accounting of this remains to be found\cite{MaMa1}.  If so, possibly the baby universe dynamics could induce couplings like those of \eqref{deltah}.

{\bf Interaction strengths and transfer rates.} 
In a description with effective interactions such as \eqref{deltah}, a key question regards the spatial dependence and form of the operator action on the BH Hilbert space of $H_{\mu\nu}$.  An important constraint is that the interactions need to be large enough to transfer sufficient information from the BH, expected to be of order one qubit per light crossing time $R$ of the BH.  This represents an $\calo(1)$ modification to the state of the Hawking radiation.  One way to achieve this\cite{SBGmodu} is if the expectation value $\langle H_{\mu\nu}(x)\rangle$ in a typical BH state is $\calo(1)$ and varies on spatiotemporal scales $\sim R$.  The question of the connection between size of couplings between subsystems and information transfer rates is a general one.  Ref.~\cite{NVU} found another way to achieve sufficient information transfer rates, with couplings exponentially small in the BH entropy, but an enhancement due to the large number of BH final states, and formulated a general conjecture about the connection between such weak couplings and transfer rates, which has been checked in simple cases\cite{GiRo}.  The former case can be called a ``strong" or ``coherent" scenario, and the latter a ``weak" or ``incoherent" scenario since there $\langle H_{\mu\nu}(x)\rangle$ is also exponentially small in the BH entropy.

{\bf Observational probes.}  There is a now widespread view that resolution of the unitarity crisis requires  new physics at horizon (or larger) scales.   This scale is now being probed by gravitational wave and very long baseline interferometric (VLBI) observations.  This coincidence of theoretical and observational developments begs further investigation.  For example, the ``nonviolent unitarization" scenarios\cite{NVU,BHQU} just described may have observable signatures\cite{GiPs,Giddings:2017jts,BHQU}.  It is also important to investigate possible observational consequences of other proposed resolutions to the crisis.

Specifically, in the strong scenario, the order unity $\langle H_{\mu\nu}(x)\rangle$ behaves like a classical perturbation of the metric, and can affect light propagation.  If the scale for these interactions extends the natural length scale $\sim R$ outside the horizon, these can alter the observed image of a BH viewed via VLBI.  Simple models of these effects were explored in \cite{GiPs}, which showed the possibility of significant time-dependent distortion of the BH shadow, and thus sufficient sensitivity to provide constraints from observation. Systematic investigation of these constraints is left for future work.

While the weak scenario does not appear to produce a similarly dramatic effect on BH images, due to smallness of $\langle H_{\mu\nu}(x)\rangle$, general arguments\cite{NVU,BHQU} indicate that it is expected to alter absorption of modes with wavelengths $\sim R$ by an amount that can be $\calo(1)$.  Since a coalescing binary produces  significant radiation at these wavelengths, such absorption can alter the binary dynamics and waveform.  The size of these modifications, and potential sensitivity of gravitational wave detectors to these, is the subject of current work.

\vskip.3in
\noindent{\bf Acknowledgements} 
This material is based upon work supported in part by the U.S. Department of Energy, Office of Science, under Award Number {DE-SC}0011702.

\mciteSetMidEndSepPunct{}{\ifmciteBstWouldAddEndPunct.\else\fi}{\relax}
\bibliographystyle{utphys}
\bibliography{BH}{}

\end{document}